\documentclass[final,5p,times,twocolumn]{elsarticle}

 \usepackage{graphicx}

\usepackage{amssymb}
\usepackage{epstopdf}
\usepackage{latexsym}

\journal{Physics Letters B}

\begin{document}

\begin{frontmatter}



\title{$\Upsilon$ Suppression in PbPb Collisions at $\sqrt {s_{NN}}$ = 2.76 TeV}


\author{Felix Nendzig and Georg Wolschin}

\address{ Institut f{\"ur} Theoretische 
Physik
der Universit{\"a}t Heidelberg, 
        Philosophenweg 16,  
        D-69120 Heidelberg, Germany, EU}

\begin{abstract}
We suggest that the combined effect of screening, gluon-induced dissociation, collisional damping, and reduced feed-down explains most of the sequential suppression of $\Upsilon(nS)$ states that has been observed in PbPb relative to $pp$ collisions at  $\sqrt{s_{NN}}$ = 2.76 TeV. The suppression is thus a clear, albeit indirect, indication for the presence of a qgp.
\end{abstract}


\end{frontmatter}



\newpage

The suppression of quarkonium states is a very promising probe for the properties of the quark-gluon plasma (QGP) that is generated in heavy-ion collisions at high relativistic energies. In the QGP the confining potential of heavy quarkonium states is screened due to the interaction of the heavy quark and the antiquark with medium partons and hence,  charmonium and bottomium states successively melt \cite{ms86} at sufficiently high temperatures $T_{{diss}}$ beyond the critical value $T_c\simeq 170$ MeV. However, additional processes such as gluon-induced dissociation, and collisional damping contribute to the suppression. Here we concentrate on such processes.


Charmonium suppression has been studied since 1986 in great detail both theoretically, and experimentally at energies reached at SPS and RHIC \cite{klusa09,kha07,eta09,ps01}, and at the LHC \cite{ma11,csi11}. The precise origin is still under investigation.
Bottomium suppression is expected to be a cleaner probe. The $\Upsilon(1S)$ ground state with invariant  mass 9.46 GeV 
is strongly bound, the threshold to $B\bar{B}$ decay is at 1.098 GeV. Its lifetime of 1.22$\cdot 10^{-20} $s is about 1.7 times as large as the one of $J/\Psi(1S)$ in elementary collisions. It melts as the last quarkonium in the QGP (depending on the potential) only at 4.10 $T_c$ \cite{won05}, whereas the $2S$ (10.02 GeV) and $3S$ (10.36 GeV) states melt at about 1.6 and 1.2 $T_c$, respectively.
Even at LHC energies the number of bottom quarks in the QGP remains small such that statistical regeneration of the $\Upsilon$ states is unimportant.

 $\Upsilon$ suppression in heavy-ion collisions has recently been observed for the first time both by the STAR experiment at RHIC \cite{hma11}, and in 2010 by the Compact Muon Solenoid (CMS) experiment at LHC \cite{chat11,cha11}. The latter includes an observation of the enhanced suppression of the $2S+3S$ relative to the $1S$ ground state. CMS data from the 2011 run shown at this conference \cite{td12} have much better statistics such that the $2S$ state can be resolved individually. 


In this work we investigate the suppression of $\Upsilon(1S), (2S), (3S)$ states at LHC energies due to  screening, gluon-induced dissociation \cite{bgw12}, collisional damping, and reduced feed-down from the $\Upsilon(2S,3S)$ and $\chi_{b}(1P)$ and $\chi_{b}(2P)$ states. Whereas gluodissociation below $T_c$ is not possible due to confinement, it does occur above $T_c$ where the color-octet state of a free quark and antiquark can propagate in the medium.
The process is relevant below the dissociation temperature $T_{{diss}}$ that is due to Debye screening, and its significance increases substantially with the rising gluon density at LHC energies.  

In the midrapidity range $|y|<2.4$ where the CMS measurement \cite{cha11} has been performed, the temperature and hence, the thermal gluon density is high, and causes a rapid dissociation in particular of the $2S$ and $3S$ states, but also of the $1S$ ground state. At larger rapidities up to the beam value of $y_{beam}=7.99$ and correspondingly small scattering angles where the valence-quark density is high \cite{mtw09}, nonthermal processes would be more important than in the midrapidity region that we are investigating here. 
Thermal gluons will also dissociate the  
 $\chi_{b}(1P)$ and $\chi_{b}(2P)$ states which partially feed the $\Upsilon(1S)$ ground state in elementary collisions \cite{aff00}.


Due to the small velocity $v\ll c$ of the quarks in the bound state, the proper equation of motion for single-particle quarkonium states is the Schr\"odinger equation, with the color-singlet $Q\bar{Q}$ quarkonium potential $V_{Q\bar{Q}}$. 
Reasonable parametrizations of the potential exist that have been tested in detailed calculations of the excited states.

In particular, the Cornell potential \cite{ei75} has string and Coulomb part 
$V_{Q\bar{Q}} = \sigma r - \alpha_{s}^s/r$,	
where $\sigma \simeq 0.192$ GeV$^2$ \cite{ja86} is the string tension, and $\alpha_{s}^s\simeq 0.37$   
the strong- coupling constant at the soft scale $m_b\alpha_s$ ($m_b\simeq 4.77$ GeV)  that accounts for the short-range gluon exchange, respectively. We shall later also refer to the coupling constants $\alpha_{s}^h\simeq 0.24$ at the hard scale $m_b$, and $\alpha_{s}^u\simeq 0.48$ at the ultra-soft scale $m_b\alpha_s^2$. 


Although the string contribution to the potential vanishes for light quarkonia in the QGP above $T_c$, it has to be considered at $T>T_c$ for heavy quarkonia that remain initially  confined and are therefore not in thermal equilibrium with the plasma. Hence we maintain the string contribution in an approximate solution of the gluodissociation problem \cite{bgw12}.

The string tension of heavy quarkonia decreases with increasing temperature $T$ in the quark-gluon medium. 
The screened potential including the imaginary part that accounts for collisional damping can be written as \cite{ja86,kms88,ber08,lai07}
\begin{eqnarray}
    \lefteqn{
V(r,T) = \sigma r_D \left[ 1 - e^{- r/r_D}\right]  -
\frac{4\alpha_{s}^s}{3} \left[ \frac{1}{r_D} +\frac{1}{r}e^{- r/r_D}\right]	}\nonumber\\&&
\qquad -i\frac{4\alpha_s^s}{3}T\int_0^\infty dz \frac{2z}{(1+z^2)^2}\left[ 1-\frac{\sin(rz/r_D)}{(rz/r_D)}\right]
\end{eqnarray}
with $r_D(T)$ the Debye radius, $ r_D^{-1} =T\,[4\pi\alpha_s^h (2 N_c + N_f)/6]^{1/2}$. 
The number of colors is $N_c=3$, the number of flavors in the QGP taken as $N_f=3$.
Because of the inverse proportionality of the minimum screening radius that permits a bound state to the heavy-quark mass \cite{ms86}, it is much more difficult to dissolve the $\Upsilon(1S)$ in the quark-gluon plasma through screening  than the $J/\psi(1S)$. 
\begin{figure}
\begin{center}
\includegraphics[width=8.0cm]{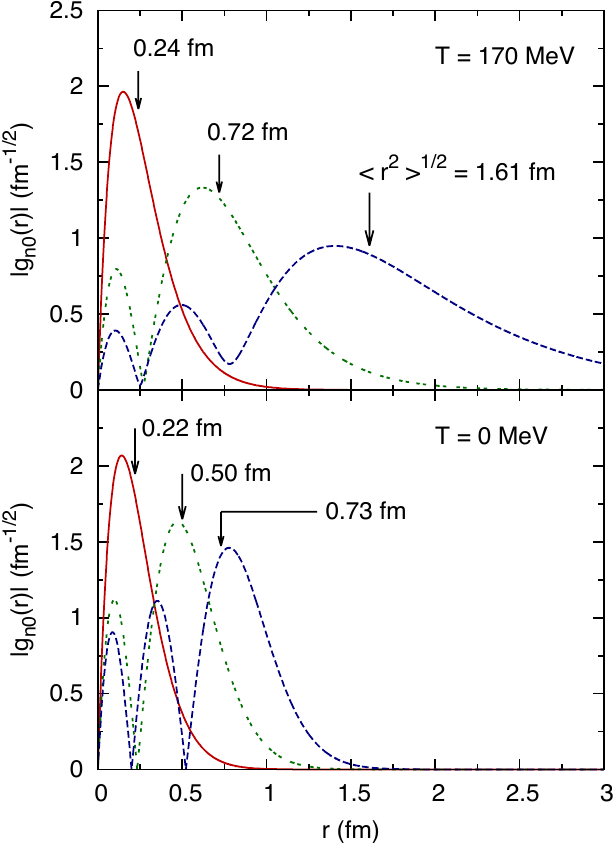}
\caption{\label{fig1} (color online) Radial wave functions of the $\Upsilon (1S), (2S), (3S)$ states (solid, dotted, dashed curves, respectively) calculated in the complex screened potential eq.(1) for temperatures $T=0$ MeV (bottom) and 170 MeV (top) with effective coupling constant $\alpha_{{eff}}\simeq (4/3)\alpha_s^s=0.49$, and string tension $\sigma=0.192$ GeV$^2$. The rms radii $<r^2>^{1/2}$ of the
$2S$ and, in particular, $3S$ state strongly dependend on temperature $T$, whereas the ground state remains nearly unchanged.} 
\end{center}
\end{figure}
We have calculated the wave functions of the individual states, as shown in Fig.~\ref{fig1} for the $1S-3S$ states at $T=0$ and 200 MeV. They are almost independent on temperature for the ground state. For the $2S$ state, there is an increase of the rms radius from $0.50$ to $0.72$ fm, whereas for the $3S$ state the rms radius increases from $0.73$ to $1.61$ fm. 

Due to the high temperature and ensuing large thermal gluon density reached at LHC energies in the midrapidity region, the most important processes next to screening that lead to a suppression of Upsilons at LHC are gluodissociation \cite{bgw12}, and collisional damping through an imaginary-valued contribution to the potential \cite{lai07,ber08,bram08,bra11} (sometimes referred to as Landau damping of the exchanged gluon \cite{str11}). 
     


Whereas gluodissociation of a specific state occurs in a limited region of the gluon energy  due to the singlet-to-octet matrix element involved, collisional damping rises with increasing gluon energy.
Hence we calculate the gluodissociation and damping cross sections for the $\Upsilon (nS)$, $\chi_{b}(1P)$, and $\chi_{b}(2P)$ states as functions of the initial impact-parameter dependent temperature in the quark-gluon plasma, and evaluate the time evolution in longitudinal and transverse space with collective velocities of 0.9c and 0.6c, respectively. Collisional damping is computed from a numerical solution of the Schr\"odinger equation for a complex potential (see also\cite{jw12}), with an imaginary part as in \cite{lai07}. Details of the calculation will be given in \cite{ngw12}.

The leading-order dissociation cross section of the $Q\bar{Q}$ states through $E1$ absorption of a single gluon had been derived by Bhanot and Peskin (BP) \cite{bp79}. 
Modifying the BP approach \cite{bgw12} to approximately account for the confining string contribution, we use the singlet wave functions computed with Eq.(1). Inserting a complete set of eigenstates 
of the adjoint (octet) Hamiltonian 
$-\Delta/m_b+\alpha_{s}^u/(6r)$ with eigenvalues $k^2/m_b$ 
to calculate the dissociation cross sections of the $\Upsilon(1S,2S,3S)$ and the $\chi_b(1P,2P)$ states \cite{nen12}, we obtain 
\begin{eqnarray} 
\sigma_{{diss}}^{nS}(E) = \frac{2 \pi^2 \alpha_s^u E}{9} \int\limits_0^\infty dk \,  \delta\left( \frac{k^2}{m_b} +\epsilon_n-E \right) 
|w^{nS}(k)|^2\quad  
	\label{improvedsigma2}
\end{eqnarray}
with the wave function overlap integral
\begin{equation}
w^{nS}(k)=\int_0^\infty dr \, r \, g_{n0}^s(r) g_{k1}^a(r) 
\end{equation}
for the singlet radial wave functions $g_{n0}^s(r)$ of the $b$ quark, and the adjoint octet wave functions $g_{k1}^a(r)$. 
The binding energy of the $nS$ state is $\epsilon_n$, and the $\delta$ function accounts for energy conservation, $k^2/m_b=E-\epsilon_n$.

For vanishing string tension $\sigma \rightarrow 0$ and the corresponding values of the binding energy $\epsilon_n$,  a pure Coulomb $1S$ wave function, and a simplification in the octet wave function, this expression reduces to the result in \cite{bp79}. 

\begin{figure}
\begin{center}
\includegraphics[width=8.6cm]{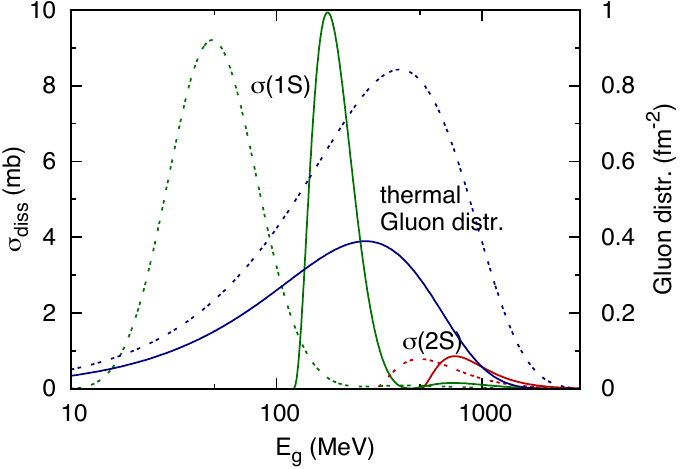}
\caption{\label{fig2} (color online) Gluodissociation cross sections $\sigma_{diss}(nS)$ in mb (lhs scale) of the  $\Upsilon (1S)$ and $\Upsilon(2S)$ states calculated using the screened wave functions calculated from the complex potential eq. (1) for temperatures $T=$ 170 (solid curves) and 250 MeV (dotted curves) as functions of the gluon energy $E_g$. 
The thermal gluon distribution (rhs scale, solid curve for $T=170$ MeV, dotted for 250 MeV)
is used to obtain the thermally averaged gluodissociation cross sections.}
\end{center}
\end{figure}

We obtain new results for the $2S$ and $3S$ states from eqs.~(2),(3). We also calculate the cross sections for the $\chi_b(1P)$ and $\chi_b(2P)$ states. The  gluodissociation cross sections resulting from eqs.~(\ref{improvedsigma2}),(3) including the effect of screening for finite string tension are shown in Fig.~\ref{fig2} for the $1S$ and $2S$ states.


     



One should be prepared to expect modifications in the cross section values of the five states from next-to-leading order (NLO) contributions \cite{sol05}, where a gluon appears in the final state in addition to the $b$ and $\bar{b}$ quarks, and hence, the phase space is larger than in leading order (LO). However, in \cite{gra01} it was shown that the quasi-free process that corresponds to NLO is less important than LO for temperatures $T>$270 MeV.

Whereas the heavy quarkonium is not in thermal equilibrium with the QGP, it is reasonable to assume that the medium itself is thermalized due to the short equilibration time of about 0.6 fm/$c$ \cite{won05}, at least in the transverse direction.
Hence, we integrate the gluodissociation cross sections for the $\Upsilon(nS)$ and $\chi_b(nP)$ states over the gluon momenta $p$, weighted with the Bose-Einstein distribution function of gluons at temperature $T$ 
to obtain the average dissociation cross sections for the $nS$ states
\begin{equation}\label{eq:avcross}
<\sigma_{{diss}}^{nS}>=\frac{g_d}{2\pi^2n_g}\int_0^\infty \sigma_{{diss}}^{nS}(E)\;\frac{p^2dp}{\exp{[E(p)/T]}-1}
\end{equation}
with $E(p)=(p^2+m_g^2)^{1/2}$, the gluon degeneracy $g_d$=16, and the gluon density as the integral over the distribution function,  $n_g=g_d T^3 \zeta(3)/\pi^2$ for $m_g=0$. Values for the thermal gluon density  at temperatures 170, 200, 300 and 400 MeV and $m_g=0$ are $n_g=$ 1.25, 2.03, 6.85 and 16.23 fm$^{-3}$, respectively.
The distribution function is shown in Fig.~\ref{fig2} (rhs scale).
\begin{figure*} 
\begin{center}
\includegraphics[width=18cm]{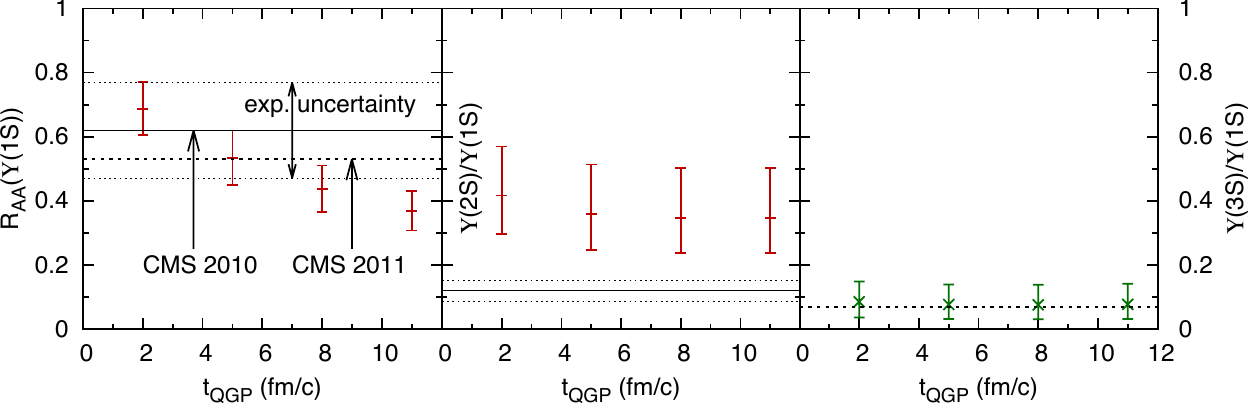}
\caption{\label{fig3} (color online) Suppression factor $R_{AA}$ for the $\Upsilon(1S)$ state (lhs), ratio
$\Upsilon(2S)/\Upsilon(1S)$ (middle), and  $\Upsilon(3S)/\Upsilon(1S)$ (rhs) calculated in the present work for 2.76 TeV PbPb minimum-bias collisions from screening, gluodissociation, collisional damping and feed-down for a $1S$ formation time  $t_F=0.1$ fm/$c$ as functions of the quark-gluon plasma lifetimes $t_{QGP}$. The corresponding CMS minimum bias results are shown from the 2010 run \cite{cha11} for $R_{AA}(1S)$  (lhs, solid line with dotted statistical and systematic uncertainties), and preliminary results from the 2011 run \cite{td12} (lhs, dashed line). The middle graph for $2S/1S$ shows our results (bars) above the preliminary CMS data from \cite{td12} (solid line, with experimental uncertainties dotted), leaving room for additional suppression mechanisms, or modifications of the formation time scales. The $3S/1S$ calculations (rhs) are compared with the preliminary CMS upper limit (dashed line \cite{td12}).  The estimated theoretical error bars account for the uncertainties in the input data that enter our calculations.}
\end{center}
\end{figure*}

The on-shell gluon energy $(p^2+m_g^2)^{1/2}$ is usually calculated assuming vanishing gluon mass $m_g=0$. The effect of a finite effective gluon mass has been investigated in \cite{bgw12}. 
Collisional damping is added as discussed in \cite{ngw12}.


The dissociation widths $\Gamma(nS,nP)$ of the $nS, nP$ states are then obtained by multiplying the average dissociation cross sections with the gluon density. The total widths are obtained by adding the corresponding damping widths. In a dynamical calculation of the fireball evolution \cite{ngw12} with a functional form
of the temperature dependence proportional to volume$^{-1/4}$ (according to the evolution of the energy density), we obtain preliminary suppression factors $\hat{R}(nS,b)$ (and analogously for the $\chi_b(nP)$ states) prior to feed-down at impact parameter $b$. 
At each impact parameter, the dissociation stops once $T_c=170$ MeV is reached. Related calculations have been performed in \cite{shk12}.
 

We then consider the subsequent radiative and hadronic feed-down cascade from the $\Upsilon(2S,3S)$ and $\chi_b$ states to the $\Upsilon(1S)$ state.
In particular, we consider $\chi_b$ populations estimated from the CDF feed-down results 
\cite{aff00}.  
With decay rates for the $nS$ states from the particle data group,
we calculate a decay cascade that matches the final populations measured by CMS for $pp$ at 2.76 TeV \cite{cha11}, and thus provides initial populations which we use for the PbPb in-medium calculation at the same energy. 

Following the consideration of screening, gluodissociation and damping of the five states, we calculate the radiative feed-down cascade in the medium for those states which have survived the strong-interaction processes at a given impact parameter $b$, to obtain the final yields in the presence of the QGP.


For an $\Upsilon(1S)$ formation time of $t_F=0.1$ fm/$c$ and a quark-gluon plasma lifetime of $t_{QGP}\simeq$ 8 fm/$c$, our result $R_{AA}(1S)\simeq 0.44$ 
is still consistent with the experimental value observed by CMS in \cite{csi11,chat11}, $R_{AA}(1S)= 0.62\pm0.11$(stat)$\pm0.10$(sys) in minimum-bias PbPb collisions, see also Fig. 3 (lhs). With these values for $\Upsilon(1S)$ formation time and qgp lifetime, the central temperature at formation time is 670 MeV.
The preliminary CMS result from the 2011 run for the ground state suppression is 0.53 \cite{td12}).
The suppression factor of the ground state may, however, be further reduced by cold nuclear matter effects such as gluon shadowing and nuclear absorption which we have not considered here. 


For the population ratio of the excited $2S$ state relative to the $1S$ ground state in minimum-bias 2.76 TeV PbPb collsions we obtain for an $\Upsilon(1S)$ formation time of $t_F=0.1$ fm/$c$ and a quark-gluon plasma lifetime of $t_{QGP}\simeq$ 8 fm/$c$ a model result of 
$\Upsilon(2S)/\Upsilon(1S)_{{PbPb}}=0.35+0.15/-0.13.$ Here the theoretical error bars arise from uncertainties in the experimental input data that we use in the calculation, such as the populations of the $\chi_b$ states.
The formation times of the excited states - which have much larger rms radii - are shorter than the one for the ground state. Here we take 0.05 fm/$c$ for the $2S$ and $1P$ states, and 0.02 fm/$c$
for the $3S$ and $2P$ states.


CMS has measured a smaller value  $\Upsilon(2S)/\Upsilon(1S)_{{PbPb}}=0.12\pm$0.03(stat)$\pm$0.01(sys) \cite{td12}, and an upper limit of $0.07$ for the ratio $3S/1S$. Hence, our result (see Fig. 3) leaves room for additional suppression mechanisms in PbPb collisions, in particular, regarding the excited  $\Upsilon(2S)$ state.

The expected physical effect, namely, rising dissociation with rising temperature, is born
out in our approach through the combination of screening, gluodissociation, collisional damping and feed-down, even though the thermally averaged gluodissociation cross sections first rise and then fall with increasing temperature for the $\Upsilon(nS)$ and $\chi_b(nP)$ states.

To conclude, we have calculated the gluodissociation, collisional damping and screening of $\Upsilon (nS)$ and $\chi_b(nP)$ states at LHC energies, plus the subsequent radiative feed-down via the $\Upsilon(2S,3S)$ and $\chi_b$ states. The weakly bound $3S$ state dissolves due to screening already at temperatures $T\gtrsim 180$ MeV which are close to the critical value. For $2S$ relative to the $1S$ state we find a substantial suppression due to screening, gluodissociation, damping and feed-down that is, however, not as large as the value reported by CMS \cite{td12}. Hence, it allows for additional suppression mechanisms of the excited state.

We obtain results for the suppression of the excited $\Upsilon$ states relative to the ground state in PbPb collsions at LHC energies with $\Upsilon(1S)$ formation time 1 fm/$c$ and QGP lifetimes of up to
11 fm/$c$. Typical central QGP temperatures have a strong time dependence. For central collisions, the initial temperature at a $\Upsilon(1S)$ formation time of 0.1 fm/$c$ is 670 MeV. 


Although screening of the strongly bound $1S$ ground state is negligible, we find that its gluodissociation is sizeable due to the strong overlap of the $1S$ gluodissociation cross section with the thermal gluon distribution, and also collisional damping is relevant. The observed suppression factor $R_{AA}(1S)\simeq 0.62$
in minimum-bias PbPb collisions \cite{cha11} (0.53 in the preliminary 2011 data \cite{td12}) is mainly due to both direct gluodissociation and damping of the $1S$ state, and to the melting and gluodissociation of the $\chi_{b}(1P)$ and $\chi_{b}(2P)$ states which partially feed the $1S$ state in $pp, p\bar{p}$ and $e^+e^-$ collisions.

The preliminary CMS data from the 2011 run that are available now \cite{td12} provide also the centrality dependence of the suppression factors for both $1S$ and $2S$ states. We shall soon compare with these data \cite{ngw12}. \\\\
\bf{Acknowledgments}

\rm
This work has been supported
by the ExtreMe Matter Institute EMMI, and IMPRS-PTFS Heidelberg.
\bibliographystyle{elsarticle-num}
\bibliography{gw_up_nt}


\end{document}